\documentclass[runningheads]{llncs}

\usepackage{graphicx}

\usepackage{tikz}
\usetikzlibrary{calc,positioning,arrows}
\usepackage{verbatim}
\usepackage{fancyvrb}
\usepackage{url}
\usepackage{xcolor}
\usepackage{todonotes}

\usepackage{amsmath}
\usepackage{amssymb}

\begin{document}

\title{Formal Methods Analysis of the\\Secure Remote Password Protocol}


\author{Alan T. Sherman\inst{1}\orcidID{0000-0003-1130-4678} \and 
Erin Lanus\inst{2}\orcidID{0000-0001-8263-0521} \and \\
Moses Liskov\inst{3}\orcidID{0000-0002-7932-0909} \and
Edward Zieglar\inst{4}\orcidID{0000-0001-5107-2125} \and \\
Richard Chang\inst{1}\orcidID{0000-0001-5278-7958} \and 
Enis Golaszewski\inst{1}\orcidID{0000-0002-0814-9956} \and \\
Ryan Wnuk-Fink\inst{1}\orcidID{0000-0002-4964-1381} \and 
Cyrus J. Bonyadi\inst{1}\orcidID{0000-0002-3686-7242} \and \\ 
Mario Yaksetig\inst{1}\orcidID{0000-0003-2175-1957} \and 
Ian Blumenfeld\inst{5}\orcidID{0000-0003-2080-9790} 
}

\authorrunning{E. Lanus et al.}

\institute{Cyber Defense Lab, University of Maryland, Baltimore County (UMBC), Baltimore, MD 21250, USA,
\email{sherman@umbc.edu} \\
\and
Virginia Tech, Arlington, VA 22309, USA,
\email{lanus@vt.edu}\\
\and
The MITRE Corporation, Burlington, MA 01720, USA,
\email{mliskov@mitre.org}
\and
National Security Agency, Fort George G. Meade, MD 20755, USA,
\email{evziegl@nsa.gov}
\and
Two Six Labs, Arlington, VA 22203, USA,
\email{ian.blumenfeld@twosixlabs.com}
}

\maketitle              

\begin{abstract}
We analyze the {\it Secure Remote Password (SRP)} protocol
for structural weaknesses using the {\it Cryptographic Protocol Shapes Analyzer (CPSA)} in the first formal analysis of SRP (specifically, Version~3).


SRP is a widely deployed
{\it Password Authenticated Key Exchange (PAKE)} protocol 
used in 1Password, iCloud Keychain, and other products.
As with many PAKE protocols, two participants 
use knowledge of a pre-shared password to authenticate each other and 
establish a session key. 
SRP aims to resist dictionary attacks, 
not store plaintext-equivalent passwords on the server, 
avoid patent infringement, and 
avoid export controls by not using encryption.
Formal analysis of SRP is challenging in part because existing tools
provide no simple way to reason about its use of the mathematical expression
``$v + g^b \mod q$''.


Modeling $v + g^b$ as encryption, we complete an exhaustive study of all possible execution sequences of SRP.  Ignoring possible algebraic attacks, this analysis detects no major structural weakness, and in particular no leakage of any secrets.  
We do uncover one notable weakness of SRP, which follows from its design constraints.  It is possible for a malicious server to fake an authentication session with a client, without the client's participation.  This action might facilitate an escalation of privilege attack, if the client has higher privileges than does the server.  
We conceived of this attack before we used CPSA
and confirmed it by generating corresponding execution shapes using CPSA.


\end{abstract}


\section{Introduction}
\label{sec:intro}


Cryptographic protocols underlie most everything that entities do in a networked computing environment,
yet, unfortunately, most protocols have never undergone any formal analysis.  
Until our work, this situation was true for
the widely deployed {\it Secure Remote Password (SRP)} 
protocol~\cite{Wu1998,Wu2000,Wu2002,GreenBlog2018_SRP}.
Given the complexity of protocols and limitations of the human mind, 
it is not feasible for experts to find all possible structural flaws in a protocol;
therefore, formal methods tools can play an important role in protocol analysis.

Protocols can fail for many reasons, including structural flaws, weak cryptography, unsatisfied hypotheses, improper configuration, inappropriate application, and implementation errors.   
We focus on structural weaknesses: fundamental logic errors, which enable an adversary to defeat a protocol's security objective or learn secret information. 

We analyze SRP for structural weaknesses 
in the first formal analysis of SRP (specifically, Version~3, known as {\it SRP-3}).  
Using the {\it Cryptographic Protocol Shapes Analyzer (CPSA)}~\cite{Liskov2016} tool
in the {\it Dolev-Yao network intruder model}~\cite{Dolev1981}, 
we model SRP-3 and examine all possible execution sequences of our model.
CPSA summarizes these executions with graphical ``shapes,'' which we interpret.

SRP is a 
{\it Password Authenticated Key Exchange (PAKE)} protocol 
used in 1Password, iCloud Keychain, and other products.
As with many PAKE protocols, two participants
use knowledge of a pre-shared password to authenticate each other and 
establish a session key. 
SRP aims to resist dictionary attacks, 
not store plaintext-equivalent passwords on the server, 
avoid patent infringement, and 
avoid export controls by not using encryption.

Formal analysis of any protocol is challenging, and analysis of SRP is particularly difficult
because of its use of the mathematical expression ``$v + g^b \mod q$''.  This expression involves both modular exponentiation and modular addition, exceeding the ability of automated protocol analysis tools to reason about modular arithmetic.
Although SRP claims to have no encryption, ironically, we overcome this difficulty by modeling the
expression as encryption, which effectively it is.

We carried out our analysis using a new virtual protocol analysis lab created at UMBC.
Embodied as a virtual machine running on the Docker utility,\footnote{\url{www.docker.com}}
this lab includes documentation, educational modules for learning about protocol analysis, and 
three protocol analysis tools: CPSA, Maude-NPA~\cite{Escobar2016,Escobar2009}, and Tamarin Prover~\cite{tamarin2017}.

Contributions of our work include: 
(1)~The first formal analysis of the SRP-3 protocol for structural weaknesses, which we carried out using the CPSA tool. Ignoring possible algebraic attacks, this analysis detects no major structural weakness, and in particular no leakage of any secrets.  
(2)~The discovery of the first attack on SRP, in which 
it is possible for a malicious server to fake an authentication session with the client, without the client's participation. This action might facilitate an escalation of privilege attack, if the client has higher privileges than does the server.

\section{Background and Previous Work}
\label{sec:back}

We briefly review formal methods for analyzing cryptographic protocols,
CPSA, PAKE protocols, and previous work on SRP.



\subsection{Formal Methods for Analyzing Cryptographic Protocols}
\label{sec:protocols}

Several tools exist for formal analysis of cryptographic protocols, including\linebreak 
CPSA~\cite{CPSA}, {\it Maude-NPA}~\cite{Escobar2016,Escobar2009}, 
the Tamarin Prover~\cite{Schmidt2012}, and ProVerif~\cite{Blanchet2015}.
Created in 2009, CPSA outputs a set of “shapes” that describe all
possible protocol executions, which can reveal undesirable execution states including ones
caused by adversarial interference.
Developed by Escobar et al. in 2009, and written in the Maude language, 
Maude-NPA works backwards from explicitly-defined attack states.
The Tamarin Prover uses a multiset-rewriting model particularly well suited for analyzing stateful protocols. 
ProVerif is an automated cryptographic protocol verifier that operates on
representations of protocols by Horn clauses. 
We choose to use CPSA because we are more familiar with that tool, have easy access to experts,
and like its intuitive graphical output.

A variety of additional tools exist to support formal reasoning, including for cryptography.
For example, created in 2009, EasyCrypt\footnote{\url{https://www.easycrypt.info/trac/#no1 }}
supports ``reasoning about relational properties of probabilistic computations with adversarial code $\ldots$ for the construction and verification of game-based cryptographic proofs.''
Cryptol~\cite{browning2010cryptol} is a domain-specific language for cryptographic primitives. Cryptol allows for the symbolic simulation of algorithms, and thus the ability to prove properties of such by hooking into various constraint (SAT/SMT) solvers.
Additionally, interactive theorem provers, such as Isabelle or Coq, have been used to analyze cryptographic functions and protocols~\cite{paulson2001relations,bartzia2014formal}. 
These tools offer the potential to verify any property expressible in their underlying logics 
(higher-order logic or dependent type theory, respectively) but sacrifice automation. 

The 1978 Needham-Schroeder~\cite{Needham1978} public-key authentication protocol dramatically illustrates
the value of formal methods analysis and limitations of expert review.   In 1995, using a protocol analysis tool, Lowe~\cite{Lowe1995} identified a subtle structural flaw in Needham-Schroeder.  This flaw had gone unnoticed for 17 years in part because Needham and Schroeder, and other security experts, had failed to consider the possibility that the intended recipient might be the adversary.  Thus, for example, if Alice authenticates to Bob, then Bob could impersonate Alice to Charlie. CPSA easily finds this unexpected possible execution sequence, outputting a suspicious execution shape.

Cryptographers sometime present a
{\it Universal Composability (UC)} proof of security~\cite{Canetti2001}, but
such proofs as typically written are long and complex and cannot be formally verified.
For example, Jarecki, Krawczyk, and Xu's~\cite{opaque_eprint2018} UC proof
of the OPAQUE protocol is in a 61-page complex paper.
By contrast, to analyze SRP-3, CPSA requires only a relatively short and easy-to-verify input that formally defines the protocol in terms of its variables, the participant roles, and 
the messages sent and received.

\subsection{Cryptographic Protocol Shapes Analyzer}
\label{sec:cpsa}

The {\it Cryptographic Protocol Shapes Analyzer (CPSA)}~\cite{CPSA,Ramsdell2012,Liskov2016} is an open-source tool for automated formal analysis of cryptographic protocols.  The tool takes as input a model of a cryptographic protocol and a set of initial assumptions called the ``{\it point of view},'' and attempts to calculate a set of minimal, essentially different executions of the protocol consistent with the assumptions. Such executions, called ``{\it shapes},'' are relatively simple to view and understand.  Executions in which something ``bad'' happens amount to illustrations of possible attacks against the protocol.  Conversely, when some property holds in all shapes, it is a property guaranteed by the protocol.

CPSA is a tool based on {\it strand space theory}~\cite{DoghmiGuttmanThayer07}, which organizes events in a partially-ordered graph.  In strand space theory, events are transmissions or receptions of messages, and sequences of events called ``{\it strands}'' capture the notion of the local viewpoint of a participant in a network.  Protocols are defined as a set of legitimate participant roles, which serve as templates for strands consistent with the protocol requirements.  

``{\it Bundles}'' are the underlying execution model, in which every reception is explained directly by a previous transmission of that exact message; a bundle of a particular protocol is a bundle in which all the strands are either (1)~generic adversary behavior such as parsing or constructing complex messages, or encrypting or decrypting with the proper keys, or (2)~behavior of participants in the protocol consistent with the protocol roles.  

CPSA reasons about bundles indirectly by analyzing ``{\it skeletons},'' which are partially-ordered sets of strands that represent only regular behavior, along with origination assumptions that stand for assumptions about secrecy and/or freshness of particular values, such as if a key is never revealed or a nonce is freshly chosen and therefore assumed unique. Some skeletons represent, more or less, the exact set of regular behavior present in some bundle consistent with the secrecy and freshness assumptions; such skeletons are called ``{\it realized}'' skeletons.  Realized skeletons are a simplified representation of actual protocol executions.  Non-realized skeletons may represent partial descriptions of actual executions, or may represent a set of conditions inconsistent with {any} actual execution~\cite{Liskov2016}.

The CPSA tool creates visualizations of skeletons as graphs in which events are shown as circles in columns where each column represents a strand.  Arrows between strands indicate necessary orderings (other than orderings within strands, or those that can be inferred transitively).  
A {\it solid arrow} represents a transmissions of some message to a reception of exactly that message.
A {\it dashed arrow} indicates a message has been altered.
The color of circles indicates what kind of event is depicted: black circles are transmissions and blue ones are receptions, and grey ones deal with state that is assumed to not be directly observable by an attacker.  For example, Figure~\ref{fig:SRP3-client1} in Section~\ref{sec:client} shows such a visualization.

\subsection{PAKE Protocols}
\label{sec:pake}


PAKE protocols evolved over time in response to new requirements and newly discovered vulnerabilities
in authentication protocols~\cite{boneh}. 
Initially, authentication over a network was carried out simply with a username and password sent in the clear. 
Unlike terminals hardwired to a computer, networks provided new and easier ways for intruders to acquire authentication credentials. 
Passively monitoring a network often harvested credentials sufficient to gain remote access to systems.
In the 1980's, {\it Kerberos}~\cite{Steiner1988} attempted to mitigate this vulnerability by no longer transmitting passwords. 
Unfortunately, the structure of Kerberos messages and the use of passwords as keys created opportunities for password guessing and dictionary attacks against the passwords, without requiring the intruder to acquire the password file directly from the server. 
Weak, user-chosen passwords simplified such attacks.

In 1992, with their {\it Encrypted Key Exchange (EKE)} protocols, 
Bellovin and Merrit~\cite{Bellovin1992} evolved PAKE protocols to address the weaknesses in user-generated passwords as keys.  In 1996, that work led Jablon~\cite{Jablon1996} to develop the
{\it Simple Password Exponential Key Exchange (SPEKE)}, which is
deployed in the ISO/IEC 11770-4 and IEEE 1363.2 standards. 
As did Kerberos, to complicate dictionary attacks, 
SPEKE incorporated random ``{\it salt}'' values into its password computations.
Attacks against the protocol in 2004~\cite{Zhang2004}, 2005~\cite{Tang2005}, and 2014~\cite{Hao2014}, 
prompted modifications to the protocol. 
Although these and similar protocols aimed to protect against the use of weak passwords for authentication, none protected the passwords from attack on the server's password file.
Access to the server's password file provided keys to authenticate as any user on the system. 


Protection of the server's authentication file became a primary new requirement that 
Wu~\cite{Wu2000,Wu1998} aimed to address with the
{\it Secure Remote Password (SRP)} protocol in 1998.
Wu addressed this requirement by not storing the password, but instead a ``{\it verifier}'' 
consisting of a modular exponentiation of a generator raised
to the power of a one-way hash function of the password. 
Improving on earlier PAKE protocols, the way SRP incorporates a random salt into
the key computation prevents the direct use of server-stored verifiers as keys.
In 2002, weaknesses discovered against SRP-3 led to a new version, {\it SRP-6}~\cite{Wu2002}.

Unfortunately, for each password, SRP publicly reveals the corresponding salt, which
facilitates pre-computation dictionary attacks on targeted passwords.  
Aware of this vulnerability, Wu nevertheless considered SRP
a significant improvement over what had come before.
Avoiding pre-computation attacks led
to new approaches including the {\it OPAQUE} protocol~\cite{opaque_eurocrypt2018,opaque_eprint2018,GreenBlog2018}.



\subsection{Previous Work}
\label{sec:srp_prev}


SRP~\cite{rfc2944,rfc5054}
is a widely used password-authenticated key-establishment protocol,
which enables two communicants to establish a secret session key 
provided the communicants already know a common password.
SRP is faster than the authenticated Diffie-Hellman key exchange protocol,
and it aims to avoid patent infringement and export control.
In this protocol, an initiator Alice (typically a client) authenticates to a responder Bob (typically a server).

In this paper, we analyze the basic version of SRP called {\it SRP-3}.
In 2002, Wu~\cite{Wu2002} proposed a variation called {\it SRP-6}, which
mitigates a two-for-one attack and decreases communication times
by allowing more flexible message orderings.




Against a passive adversary, SRP-3 seems to be as secure as
the Diffie-Hellman problem~\cite{Diffie1978,Maurer2000,GreenBlog2018_SRP}. It remains possible, however,
that a passive adversary can acquire information from eavesdropping
without solving the Diffie-Hellman problem.
Against an active adversary, the security of SRP-3 remains unproven.

Wu~\cite{Wu2000} claims to prove a reduction from the Diffie-Hellman problem to breaking SRP-3 against a passive adversary, but his proof is incorrect: his reduction assumes the adversary
knows the password, which a passive adversary would not know.\footnote{Wu incorrectly states the direction of his reduction, but his reduction actually proceeds in the correct direction.}
We are not aware of any other previous effort to analyze the SRP protocol.

Wilson et al.~\cite{Wilson1998} survey authenticated Diffie-Helman key agreement protocols.
Adrian et al.~\cite{Adrian2015} analyze how such protocols can fail in practice.
Schmidt et al.~\cite{Schmidt2012} present automated analysis of Diffie-Helman protocols.

As an example of formal analysis of a protocol using CPSA, we note:
In 2009, Ramsdell et al.~\cite{Ramsdell2012} analyzed the CAVES attestation protocol using CPSA,
producing shapes that prove desirable authentication and confidentiality properties. 
The tool successfully analyzed the protocol despite the presence of 
hash functions and auxiliary long-term keys.
As another example, which illustrates the utility of service roles,  
see Lanus and Zieglar~\cite{Lanus_JIW_2017}

\section{The Secure Remote Password Protocol}
\label{sec:srp}

Figure~\ref{fig:SRP-3_ladder} summarizes how SRP-3 works, during which
Alice and Bob establish a secret {\it session key}~$K$, leveraging 
a {\it password}~$P$ known to Alice and Bob. 

In SRP-3, all math is performed in some prime-order group
${\mathbb Z}_q$, where $q$ is a large prime integer.
Let $g$ be a generator for this group.  The protocol uses 
a hash function~$h$.  For brevity, for any $x \in {\mathbb Z}_q$, we shall write
$g^x$ to mean $g^x \bmod q$.

Before executing the protocol, Alice must register her {password}~$P$ with Bob.  
Bob stores the values $(s, v)$ indexed by ``Alice'',
where $s$ is a random {\it salt}, \hbox{$x = h(s,P)$}
is the salted hash value of Alice's password, and
$v = g^x$ is a non-sensitive ``{\it verifier}'' 
derived from~$P$, which does not reveal $x$ or~$P$.


SRP-3 works in two phases:  I.~Key Establishment and II.~Key Verification.
The protocol establishes a new session key $K$ known to Alice and Bob,
which they can use, for example, as a symmetric encryption key.
Phase~I works as follows.

\begin{figure}
\centering
\begin{tikzpicture}

    \draw[red, solid] (-6.5,0.5) rectangle (-2.5,-0.5); 
    \node at (-4.5,0) {$P, s, x=h(s,P), v=g^x$};
    \draw[red, dashed] (2.5,0.5) rectangle (3.5,-0.5); 
    \node at (3,0) {$s,v$};
    
    \draw (-2,-0.3) -- (-2,-4) (2,-0.3) -- (2,-4); 
    \node at (-2,0) {\textbf{\thinspace\thinspace Client}};
    \node at (2,0) {\textbf{Server \thinspace\thinspace}};
    \draw[-stealth] (-2,-1) -- node[midway,above] {client} (2,-1);
    \draw[stealth-] (-2,-2) -- node[midway,above] {$s$} (2,-2);
    \draw[red, solid] (-3.5,-2.1) rectangle (-2.5,-3); 
    \node at (-3,-2.5) {$a, g^a$};
    \draw[red, dashed] (2.5,-2.1) rectangle (3.5,-3); 
    \node at (3,-2.5) {$g^a$};
    \draw[-stealth] (-2,-3) -- node[midway,above] {$g^a$} (2,-3);
    \draw[red, dashed] (-3.5,-3.1) rectangle (-2.5,-4); 
    \node at (-3,-3.5) {$g^b, u$};
    \draw[red, solid] (2.5,-3.1) rectangle (3.5,-4); 
    \node at (3,-3.5) {$b, u$};
    \draw[stealth-] (-2,-4) -- node[midway,above] {$v + g^b$,$u$} (2,-4);

    \node at (-4.15, -4.5) {$K = h((v+g^b-v)^{a+ux})$};
    \node at (-4.5, -5) {$= h((g^b)^{a+ux})$};
    \node at (-4.5, -5.5) {$= h(g^{b(a+ux)})$};
    
    \node at (3.65,-4.5) {$K = h((g^a(g^x)^u)^b)$};
    \node at (3.75,-5) {$= h((g^ag^{ux})^b)$};
    \node at (3.75,-5.5) {$= h((g^{a+ux})^b)$};
    \node at (3.75, -6) {$= h(g^{b(a+ux)})$};
    
    \draw (-2,-4) -- (-2,-8) (2,-4) -- (2,-8); 
    \draw[-stealth] (-2,-7) -- node[midway,above] {$h(g^a,v+g^b,K)$} (2,-7);
    \draw[stealth-] (-2,-8) -- node[midway,above] {$h(g^a,h(g^a,v+g^b,K),K)$} (2,-8);
    
\end{tikzpicture}
       \caption{Protocol diagram for SRP-3, which comprises three phases: setup, key exchange, and key verification. During key exchange, the server transmits to the client the expression $v + g^b \mod q$, which we cannot directly model in CPSA. Variables in dashed boxes denote values received.} 
       \label{fig:SRP-3_ladder}
\end{figure}
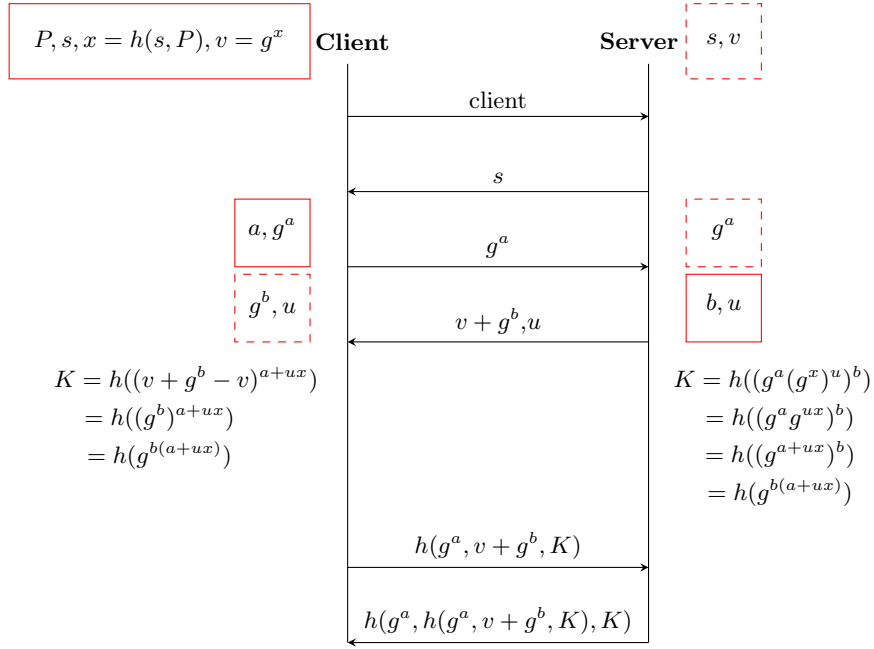{}
   
   

\begin{enumerate}
    \item Alice sends her identity ``Alice'' to Bob.
	
    \item Bob receives Alice's identity and looks up Alice's salt~$s$ and 
	stored verifier $v = g^x$, where $x = h(s, P)$.
	Bob sends Alice her salt~$s$.
	
    \item Alice receives $s$, calculates $x = h(s,P)$, and generates a random secret nonce~$a$. 
	Alice calculates and sends $g^a$ to Bob.
	
    \item Bob receives $g^a$ and generates a random secret nonce~$b$ and a random scrambling parameter~$u$.
	Bob calculates and sends $v + g^b$ to Alice, together with~$u$.
	
    \item Each party calculates the session key~$K$ as the hash of a common value, which each party computes differently. 
    Alice calculates $K = h((v + g^b) - g^x)^{a + ux}$ 
	and Bob calculates 
	$K = h(g^{a} g^{ux})^b$. 
\end{enumerate}

Thus, in Phase~I, Alice and Bob establish a common session key~$K$.
In Phase~II, Alice and Bob verify that they have the same session key.
Phase II works as follows.

\begin{enumerate}

    \item Alice computes 
    $M_1 = h(g^a, v + g^b, K)$ 
    and sends $M_1$ to Bob. 
	Bob verifies the received value by recomputing 
	$M_1 = h(g^a, v + g^b, K)$.
	
    \item Bob computes 
    $M_2 = h(g^a, M_1, K)$ 
    and sends it to Alice.
	Alice verifies the received value by recomputing 
	$M_2 = h(g^a, M_1, K)$.
	
    \item If and only if these two verifications succeed, the session key $K$ is verified.
\end{enumerate}

\section{Modeling SRP-3 in CPSA}
\label{sec:modeling-main}

Using CPSA, we analyze SRP-3 in the Dolev-Yao network intruder model in two steps: 
In this section, we model SRP-3 in CPSA; 
in the next section, we interpret shapes produced by our model.
Appendix~\ref{sec:cpsa-code} lists important snippets of our CPSA sourcecode.

\subsection{Challenges to Modeling SRP-3 in CPSA}
\label{sec:challenges}

CPSA provides two algebras to express protocols: basic and Diffie-Hellman.
The basic crypto algebra includes functions that support modeling of pairings, decomposing a pair into components, hashing, encrypting by symmetric and asymmetric keys, decrypting by keys, 
returning the ``inverse of a key'' (a key that can be used to decrypt),
and returning a key associated with a name or pair of names.
CPSA does not support arithmetic operations. 
The Diffie-Hellman algebra extends the basic crypto algebra by 
providing ``{\it sorts}'' (variable types) that represent exponents and bases, as well as functions for a standard generator $g$, a multiplicative identity for the group, exponentiation, and multiplication of exponents. 

SRP-3 is challenging to model in CPSA because CPSA does not support any of the following
computations:
addition of bases when the server sends $v + g^b$,
subtraction of bases when the client computes $(v + g^b) - v$,
and addition of exponents (i.e., multiplication of bases) when the client computes the key.
CPSA handles only multiplication of exponents, and cannot be easily modified to handle these additional algebraic operations, because 
CPSA makes use of general unifications in its class of messages,
and a full decision procedure in the theory of rings is 
undecidable~\cite{church1936unsolvable}.


\subsection{Our Model of SRP-3}
\label{sec:modeling}

We model SRP-3 by defining variables, messages, and associated roles.
Critical modeling decisions are
how to represent the problematic expression $v+g^b$,
how to deal with multiplication of bases,
and how to handle the initialization phase.
Figure~\ref{fig:SRP-diagram-CPSA} shows the SRP-3 protocol diagram
as we modeled SRP-3 in CPSA.

There are two legitimate protocol participants, which we model 
by the {client} and {server} roles (see Figure~\ref{fig:model_SRP-3}).  
We organize each of these roles into two phases: initialization and main.
The initialization phase establishes and shares the password,  and it
establishes the salt and verifier in the long-term memory of the server.

We model the problematic expression $v+g^b$ as $\{|g^b|\}_v$, 
which is the encryption of $g^b$ using $v$ as a symmetric key.  
Thus, knowing $g^b$ requires knowledge of $v$.

The other problematic expressions occur in the calculation of the key.  The key $K$ is supposed to be equal to $(g^b)^{a+ux}$.  Here, each party calculates this value by calculating $g^{ab}$ and $g^{bux}$ and multiplying them together.  The client can calculate these values from $g^b$ 
by raising $g^b$ to the $a$ power and to the $ux$ power.  The server calculates these values by raising $g^a$ to the $b$ power, and by raising $g^x = v$ to the $bu$ power.  

We emulate the multiplication of these base values by hashing them; since both parties can calculate the two factors, each can calculate the hash of the two factors.  Thus, we represent the key $K$ as $K = h(g^{ab}, g^{bux})$, where $h$ stands for cryptographic hashing.

\begin{figure}
\centering
       \begin{tikzpicture}
   
    \node at (-2.25,4) {\textbf{client-init}};
    \node at (2.25,4) {\textbf{server-init}};
    \draw (-2.25,3.7) -- (-2.25,1) (2.25,3.7) -- (2.25,1); 
    \filldraw[color=red!60, fill=red!5, solid](-4.75,3) circle (1.25);
    \node at (-4.75,3.5) {``client state,''};
    \node at (-4.75,3) {$s,x$};
    \node at (-4.75,2.5) {client, server};
    \draw[stealth-] (-3.5,3) -- node[midway,above] {init} (-2.25,3); 
    \draw[-stealth] (-2.25,2) -- node[midway, above] {$\{|``Enroll,'' s, g^x|\}_{client-server}$} (2.25,2); 
    \filldraw[color=red!60, fill=red!5, solid](4.75,1) circle (1.25);
    \node at (4.75,1.5) {``server record,''};
    \node at (4.75,1) {$s,v=g^x$};
    \node at (4.75,0.5) {client, server};
    \draw[-stealth] (2.25,1) -- node[midway,above] {init} (3.5,1); 

    \draw (-2.25,-0.3) -- (-2.25,-4) (2.25,-0.3) -- (2.25,-4); 
    \node at (-2.25,0) {\textbf{client}};
    \node at (2.25,0) {\textbf{server}};
    \draw[-stealth] (-2.25,-1) -- node[midway,above] {client} (2.25,-1);
    \draw[dashed] (2.25,-1.5) -- node[midway,above] {obsv} (4.75,-1.5); 
    \draw[dashed,-stealth] (4.75,-1.5) -- (4.75,-0.5); 
    \draw[stealth-] (-2.25,-2) -- node[midway,above] {$s$} (2.25,-2);
    \draw[dashed] (-2.25,-2.5) -- (-4.75,-2.5); 
    \draw[dashed,-stealth] (-4.75,-2.5) -- (-4.75,1.5); 
    \filldraw[white] (-4,-2.1) rectangle (-3,-3); 
    \draw[red, solid] (-4,-2.1) rectangle (-3,-3); 
    \node at (-3.5,-2.5) {$a, g^a$};
    \draw[red, dashed] (3,-2.1) rectangle (4,-3); 
    \node at (3.5,-2.5) {$g^a$};
    \draw[-stealth] (-2.25,-3) -- node[midway,above] {$g^a$} (2.25,-3);
    \draw[red, dashed] (-4,-3.1) rectangle (-3,-4); 
    \node at (-3.5,-3.5) {$g^b, u$};
    \draw[red, solid] (3,-3.1) rectangle (4,-4); 
    \node at (3.5,-3.5) {$b, g^b, u$};
    \draw[stealth-] (-2.25,-4) -- node[midway,above] {$\{|g^b|\}_v,u$} (2.25,-4);

    \node at (-4, -5) {$K = h((g^{b})^a,(g^{b})^{ux})$};
    \node at (4,-5) {$K = h((g^{a})^b,(g^{x})^{ub})$};
    
    \draw (-2.25,-4) -- (-2.25,-7) (2.25,-4) -- (2.25,-7); 
    \draw[-stealth] (-2.25,-6) -- node[midway,above] {$h(g^a,\{|g^b|\}_v,K)$} (2.25,-6);
    \draw[stealth-] (-2.25,-7) -- node[midway,above] {$h(g^a,h(g^a,\{|g^b|\}_v,K),K)$} (2.25,-7);
    
\end{tikzpicture}
       \caption{Protocol diagram for SRP-3, as we modeled it in CPSA. We introduce two service roles, {client-init} and {server-init}, that handle the setup phase by instantiating values for $s$, $x$, and $v = g^x$, and by making these values available to the legitimate client and server. We model the computation $v+g^b$ as an encryption of $g^b$ under key~$v$.  Variables in dashed boxes denote values received.}
       \label{fig:SRP-diagram-CPSA}
\end{figure}
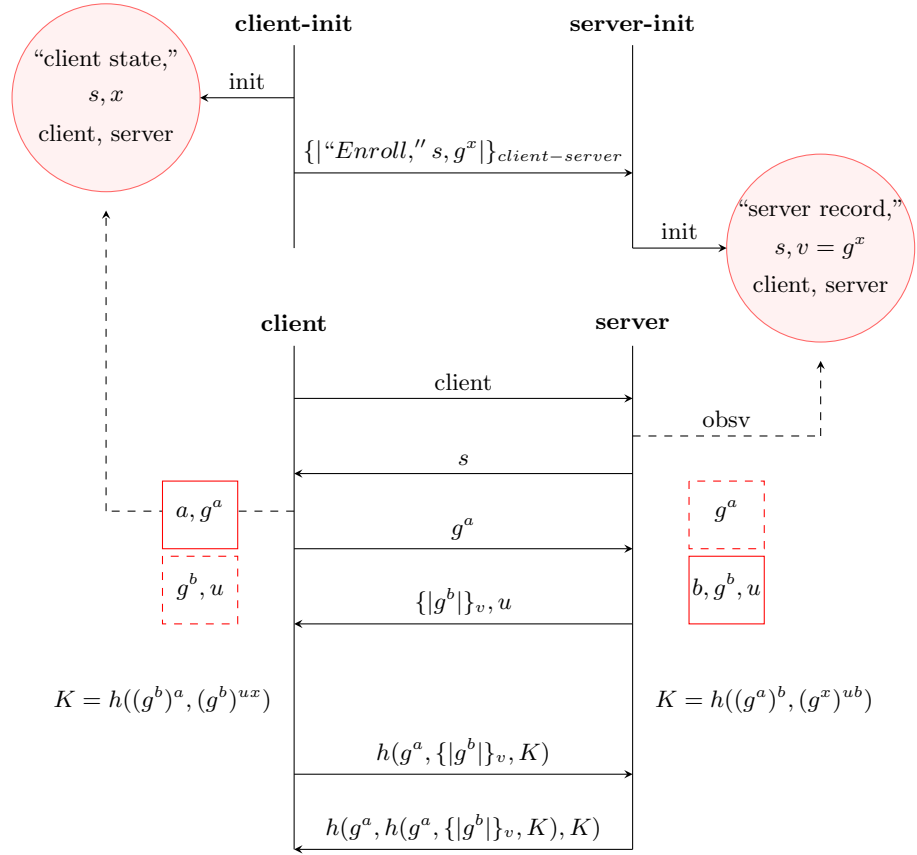{}

Finally, we explain how we model the initialization phase, and in particular, how
the client communicates their salt and verifier to the server.
In the beginning of the client and server roles, one could exchange the salt and verifier as
a message.  This strategy, however, would
prevent CPSA from exploring scenarios in which the same client or server
conducts multiple executions of the protocol 
using the same password information exchanged during initialization. 
Instead, we use ``{\it service roles},'' which provide  a  function or service to one or more participant roles. Our service roles exchange values across a secure channel.  These values persist 
in state that can be accessed only by instances of the appropriate main-phase roles. 

Specifically, the client-init service role initializes a state record with the value 
\{``client state'', $s,x$, client, server\} (see Figure~\ref{fig:model_SRP-3}).
The ``client state'' string literal serves the function of a label,
enabling us to write client roles to observe state that begins with that string. 
We store the salt and password hash because each client role directly uses these values.
The names of the client and server help to link the state to the correct client-server pair. 

After initializing its state, the client-init role sends a string literal ``Enroll'',
together with the salt and verifier. 
The client-init role encrypts this
message using a long-term key known by the particular client and server. 
The server-init role receives this message and initializes the server's state 
by storing a string literal ``server record'', the salt and verifier, and
the names of the client and server. 

To prevent CPSA from instantiating an unlimited number of server-init and client-init roles, 
we add a rule that disregards any executions in which there is more than one instance of the server-init role for a specific client / server pair, see Figure~\ref{fig:model_SRP-3_rule}.

The model above is sufficient to verify most of the security properties of SRP, but cannot verify the property that compromise of the server's authentication database cannot be used directly to compromise the security and gain immediate access to the server. The reason is that if SRP meets its security goals, the verifier $v$ is not leaked to the adversary by the protocol. 
Therefore, to test whether or not access to $v$ allows the adversary to impersonate a client to the server, we need to use a model 
in which the server-init role is modified to transmit the verifier it receives for a client. 
This model provides the adversary with access to $v$ that they cannot obtain from SRP. 
For this property, it is sufficient to test only the server's point of view. 
Compromise of a server's authentication database would allow anyone to impersonate a server to the client and is not a property that SRP was designed to prevent.

\section{Interpreting Shapes from the SRP-3 Model}
\label{sec:interpretation}

We generate and interpret shapes showing executions of our model of SRP-3
under various assumptions from the perspectives of various roles.
Specifically, we define skeletons that provide the perspectives of
an honest client and an honest server, respectively 
(see Figures~\ref{fig:model_SRP-3_client_skeleton} and~\ref{fig:model_SRP-3_server_skeleton}).
We also define ``{\it listeners}'' to detect possible 
leaked values of the password hash~$x$ or verifier~$v$
(see Figures~\ref{fig:model_SRP-3_listener_x} and~\ref{fig:model_SRP-3_listener_v}).
Finally, we investigate if an adversary directly using a compromised verifier could authenticate
as a client (see Figure~\ref{fig:SRP3-leak}).
CPSA completed its search, generating all possible shapes for each point of view 
(see~\cite{Liskov2011} for an explanation). 

Figures~\ref{fig:SRP3-client1}--\ref{fig:SRP3-leak} 
display selected shapes that highlight our main findings.
These shapes show that, when the client and server are honest,
there is no attack against our model of SRP-3:
the only way the protocol completes is between a client and a server. 
Similarly, CPSA found no leakage of~$x$ or~$v$.
CPSA also found that an adversary directly using a compromised
verifier cannot authenticate as a client without access to internal values of the server.

\subsection{Client Point of View}
\label{sec:client}

\begin{figure}
	\centering
	\def\svgscale{.3}
	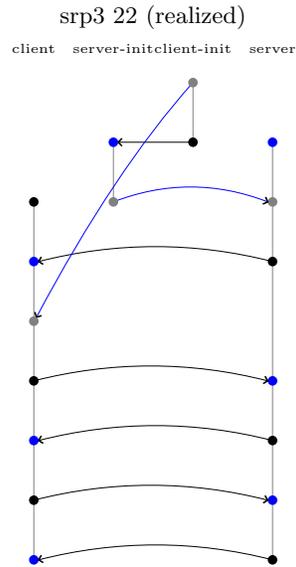
	\caption{Shape showing an execution of SRP-3 from the client's perspective. This graphical output from CPSA reveals expected behavior.}
    \label{fig:SRP3-client1}
\end{figure}

\begin{figure}
	\centering
	\def\svgscale{.3}
	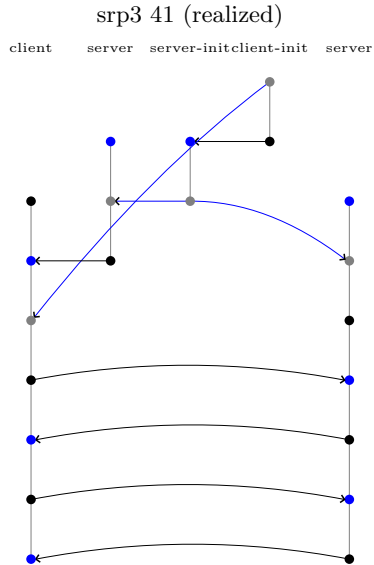
	\caption{Shape showing an execution of SRP-3 from the client's perspective, with an additional run of the server. This graphical output from CPSA reveals both server roles accessing the same state, causing them to behave like two instances of the same server. The client can begin the protocol with one instance of the server, then complete it with the other.  This unexpected shape does not
	constitute an attack.}
    \label{fig:SRP3-client2}
\end{figure}

Figures~\ref{fig:SRP3-client1} and~\ref{fig:SRP3-client2} show the two shapes
generated from the perspective of an honest client. 
The first shape is what we had expected.
One added client-init strand provides state needed for the client to access password information, and 
one added server-init strand provides password information to the server strand.
The solid lines in the shape prove that the messages must come from the expected parties, and the shape closely reflects the protocol diagram for our model. 

The second shape explores the possibility that the adversary could replay the client's initial message to the server resulting in the server beginning two protocol runs with the client. We are able to verify that it is the same server by observing that the server variables in both strands are instantiated with the same value. Only one of the server strands is able to complete, because the messages between the two runs of the protocol cannot be confused. The shape indicates that there is not any way for the adversary to take advantage of initiating multiple runs of the protocol with the server.

\subsection{Server Point of View}  
\label{sec:server}

\begin{figure}
	\centering
	\def\svgscale{.3}
	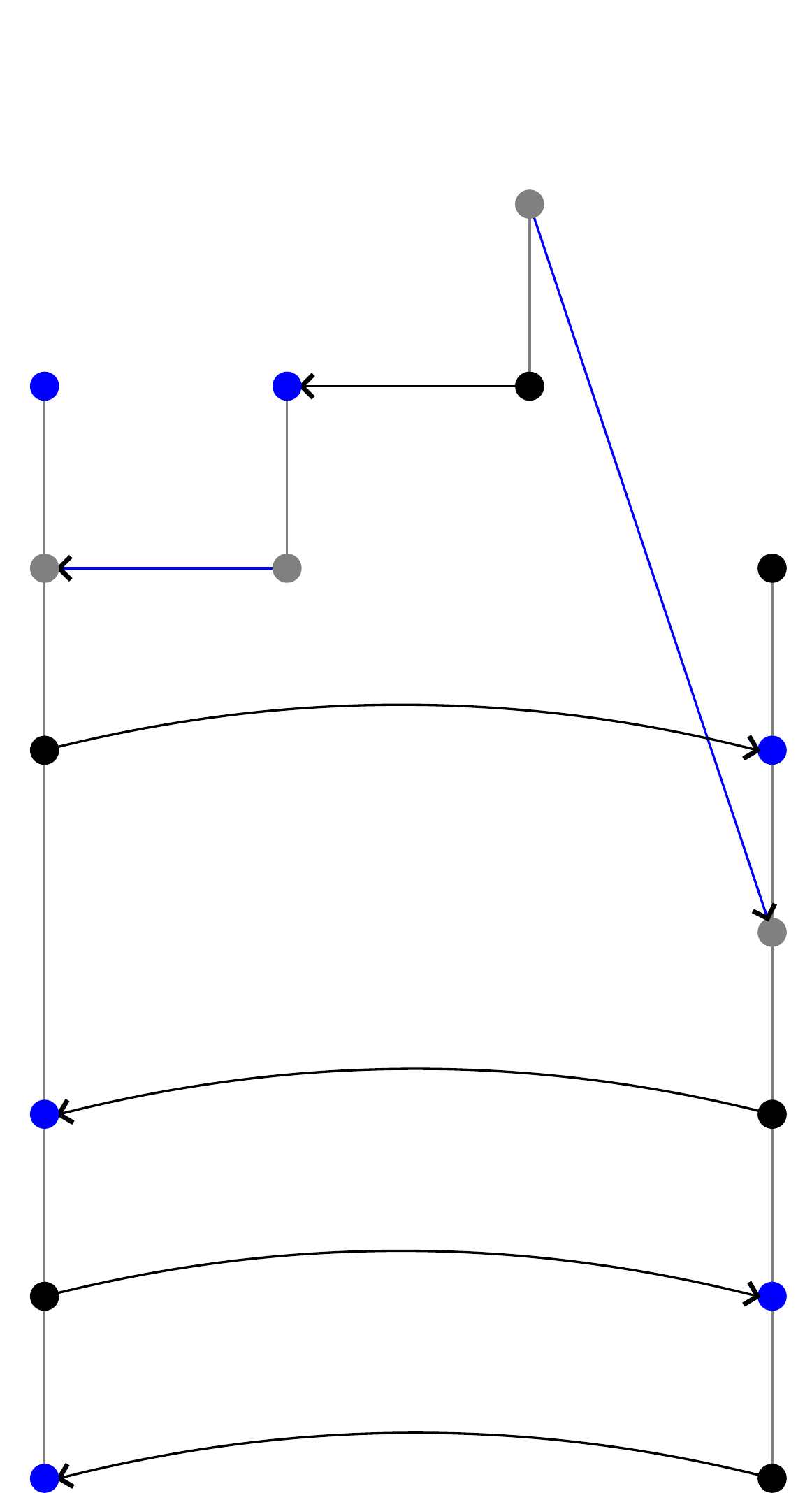
	\caption{Shape showing an execution of SRP-3 from the server’s perspective. This graph-ical output from CPSA reveals expected behavior.}
    \label{fig:SRP3-server1}
\end{figure}

Figure~\ref{fig:SRP3-server1} shows the first of two shapes generated from the perspective 
of an honest server.
As happens for the client, two shapes result. 
The first shape is similar to the protocol diagram for our model and is what we had expected.
A client is needed to complete the protocol, as are the service roles server-init and client-init. The second shape indicates a replay of the client's initial message resulting in two server strands with the same server as indicated in the strands' instantiated variables. As with the additional shape in the client's view, only one of the server's strands is able to complete, indicating that there is no attack against the protocol from the server point of view.

\subsection{Privacy Properties}
\label{sec:listeners}

It is important that the password hash~$x=h(s,P)$ and
the verifier~$v=g^x$ remain secret.  To determine whether a network adversary can observe
either of these values in our model of SRP-3, we define two input skeletons to test these privacy properties, one for~$x$ and one
for~$v$ (see Figures~\ref{fig:model_SRP-3_listener_x} and~\ref{fig:model_SRP-3_listener_v}).
Because the client knows~$x$, we add the listener for~$x$ to the client point of view.
Similarly, because the server knows~$v$, we add the listener for~$v$ to the server point of view.  Listeners in CPSA represent a test that a value can be found by the adversary.

For each if these skeletons, we ran CPSA.  In each case, CPSA returned an empty tree, meaning that there is no way to realize the skeleton as a shape, which means that no such attack is possible in our model.  In each case, CPSA ran to completion, indicating that it explored all possible shapes for the model.

\subsection{Leaked Verifiers}
\label{sec:leaks}

\begin{figure}
	\centering
	\def\svgscale{.3}
	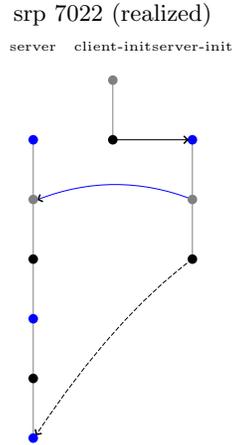
	\caption{Shape showing an execution of SRP-3 from the server’s perspective when the verifier is leaked to the adversary and $u=b$. This shape indicates an attack where the adversary completes an authentication with the server impersonating a client.}
    \label{fig:SRP3-leak}
\end{figure}

CPSA analysis of listeners for~$v$ confirms that the SRP protocol does not leak the verifier~$v$. Therefore, to analyze the protocol when the adversary has access to~$v$, we modified server-init to leak the verifier to the adversary.  In the presence of this variant of the server-init role, CPSA discovered two main shapes: one is the ordinary server point of view (Figure~\ref{fig:SRP3-server1});
the other shows that the adversary is able to impersonate a client 
if the verifier has indeed leaked (Figure~\ref{fig:SRP3-leak}).

The situation is more subtle.  The adversary is able to impersonate the client only if they
know both~$v$ and~$b$, as an adversary might learn if the adversary comprised the server. 
Initially, in our model of SRP-3, we did not require that~$b$ and~$u$
be distinct, only that they be uniquely generated.  CPSA found the impersonation attack in part because
CPSA deduced that the adversary could learn~$b$ if $b=u$, since SRP-3 reveals~$u$.
Subsequently, we added an additional assumption that $b \neq u$, when CPSA discovered
only the the expected shapes.
This fact validates the assertion that SRP is secure from an adversary directly using the verifier to authenticate as a client without access to internal values of the server.

\section{A Malicious Server Attack against SRP}
\label{sec:attack}

Our analysis in Section~\ref{sec:interpretation} assumes that legitimate participants of SRP-3 are honest, meaning they will execute the protocol faithfully.  In this section, we explore an attack on SRP-3 in which
the server is compromised.  For example, an adversary might corrupt the server to run a malicious process.  
In this attack, the malicious server authenticates to itself, pretending to be a particular client,
without the client's involvement.  A possible goal of this attack might be for the malicious server to escalate its privileges to those of the client, which might be higher than those of the server.

To analyze this attack, we define a malicious server role, which we call {\it malserver}
(see Figure~\ref{fig:model_malserver}).  We provide to malserver 
only the information that an honest server would have access to by observing the state initialized by a server-init role.  Consequently, malserver must compute the key using the same method as 
carried out by an honest server. Malserver also acts like a client, initiating the protocol and sending  messages consistent with those from the client role.
Figure~\ref{fig:model_malserver} also defines an associated skeleton, which enables CPSA to compute
a strand of the malserver role.



\begin{figure}
	\centering
	\def\svgscale{.3}
	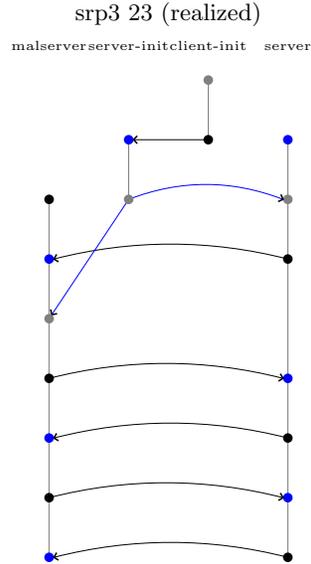
	\caption{Shape showing an execution of SRP-3 from the perspective of a malicious server. This graphical output from CPSA reveals that a malicious server can deliberately act like a client and can authenticate the client with a legitimate instance of itself. 
	Using only information available to the server role, the malicious server can thereby potentially inherit a client's higher privileges.}
    \label{fig:SRP3-malserver1}
\end{figure}

Figure~\ref{fig:SRP3-malserver1} shows the first of two shapes produced by CPSA from the malserver skeleton.  
As for honest participants, CPSA also produced a second shape that shows the protocol can be started and completed with two different honest server roles on the same machine.
Figure~\ref{fig:SRP3-malserver1} shows the malserver initiating the protocol by sending the client's name and proceeding to interact with the server as though it were the client, all the way through to the key verification messages. For executions with a legitimate client, CPSA adds client-init and server-init strands, as a result of the setup phase in which a client sends 
name, salt, and verifier to the server.  Here, however, there is no client strand. 
The server sends the final black node on its strand only after the server verifies the hash provided by the malserver strand, indicating that the server believes it is communicating with the specified client.

The attack is possible because the malserver role is operating on the server it is attacking 
(the server and malserver variables are equal) and has access to the server's internal values, as we discuss in the analysis of the leaked verifiers. 
Even though this attack is not a part of the Dolev-Yao model that CPSA uses,
by creating a special malserver role outside of the normal protocol roles, 
we were able to coax CPSA to explore the attack.
 

\section{Discussion}
\label{sec:discussion}


We briefly discuss two limitations of our work: 
one arising from our modeling of the ``problematic expression'' $v + g^b$ as encryption,
the other arising from our choice of CPSA's point of view (see Section~\ref{sec:cpsa}).

Modeling the problematic expression as encryption enabled CPSA to carry out its work.  
A consequence of this crucial decision, however, is that we analyzed a slight variation of SRP-3 that might be stronger than SRP-3. By abstracting these algebraic operations as strong encryption, our analysis cannot find possible ``algebraic attacks'' that might take advantage of detailed algebraic relationships. We are not aware of any such attacks on SRP-3 and do not suspect that they exist, but we cannot exclude their possible existence. The consequences of this crucial modeling decision are similar to those from the common practice of modeling a particular encryption function as a strong encryption function, 
which excludes the possibility of finding attacks that exploit possible weaknesses in the
particular encryption function.


CPSA exhaustively explores possible executions of a protocol from a specified point of view and set of assumptions.  Such analysis holds only when those assumptions are satisfied for that point of view. For example, initially, CPSA did not find the malicious server attack 
described in Section~\ref{sec:attack}. CPSA did not find this attack because the adversary requires access to variables~$v$ and~$b$, that are not available through the messages exchanged and the assumptions of the model. We were able to show that SRP-3 does not leak those values. The model also failed to verify SRP-3's property that access to the state variable~$v$ by the adversary would not allow the adversary to impersonate a client directly. 
To verify that property would require a model that made~$v$ available to the adversary.


Subsequently, we explored two models to investigate possible impersonation attacks.
One model gave the adversary~$v$; the other model gave the adversary~$v$ and~$b$.
With these models, CPSA showed that the adversary can impersonate the client if they
know~$v$ and~$b$, but not if they know only~$v$ (see Section~\ref{sec:interpretation}).

Different assumptions and points of view can influence analyses.
All formal method tools explore properties only within a specified scope and
do not find attacks outside that scope.  Although CPSA did not initially 
discover the malicious server attack, we were able to enlarge 
CPSA's scope of search to find it.  It is possible, however, that there
might be additional attacks outside our scope of search.



\section{Conclusion}
\label{sec:conclusion}


Using CPSA, we formally analyzed the SRP-3 protocol in the Dolev-Yao network intruder model and found it free of major structural weakness.   We did find a weakness that a malicious server can fake an authentication session with a client without the client's participation, which might lead to an escalation of privilege attack.

Limitations of our analysis stem in part from our cryptographic modeling.
CPSA will not find attacks that exploit weak cryptography, 
and our use of CPSA will not find any algebraic attacks.
As all tool users must, we trust the correctness of CPSA and its execution.  
Our results do not speak to a variety of other potential issues, including 
possible implementation and configuration errors when using SRP-3, and inappropriate applications of it.

Open problems include formal analysis of other PAKE protocols~\cite{Haase2018}, including the 
recent OPAQUE protocol~\cite{opaque_eurocrypt2018,opaque_eprint2018,GreenBlog2018}, which, unlike SRP, 
tries to resist precomputation attacks by not revealing the salt values used by the server.  
OPAQUE is the most promising new protocol possibly to replace SRP.
Because quantum computers can compute discrete logarithms in polynomial time, 
it would be useful to study and develop post-quantum PAKE protocols~\cite{Jintai2017} that can resist
quantum attack.

We hope that our work, as facilitated by the virtual protocol analysis lab created at UMBC, 
will help raise the expectation of due diligence to include formal analysis
when designing, standardizing, adopting, and evaluating cryptographic protocols.

\section*{Acknowledgments}


Thanks to John Ramsdell (MITRE) and other participants at the Protocol eXchange for fruitful interactions. 
This research was supported in part by the U.S. Department of Defense under CySP Capacity grants H98230-17-1-0387 and H98230-18-1-0321.
Sherman, Golaszewski, Wnuk-Fink, Bonyadi, 
and the UMBC Cyber Defense Lab  were supported also in part by
the National Science Foundation under SFS grant DGE-1753681.



\bibliographystyle{splncs04}
\bibliography{SRP-HCSS-bib}
 

\bigskip \noindent
{\it Submitted to Springer LNCS on February 28, 2020.}
\bigskip

\appendix
\section{CPSA Sourcecode}
\label{sec:cpsa-code}

We list critical snippets of CPSA sourcecode that we used to define SRP-3 and carry out our analysis.


\clearpage

\begin{figure}  
    \input{figures/input-for-srp3}
    \caption{Modeling of SRP-3 in CPSA. We define four roles: {client-init}, {server-init}, {client}, and {server}. The {client-init} and {server-init} roles are service roles that initialize common values between the {client} and {server} roles.}
    \label{fig:model_SRP-3}
\end{figure}


\begin{figure}
    \input{figures/added-rules}
    \caption{Rule added to SRP-3 to prevent CPSA from instantiating an unlimited number of {server-init} roles, which would prevent CPSA from terminating.}
    \label{fig:model_SRP-3_rule}
\end{figure}


\begin{figure}
    \input{figures/client-skeleton}
    \caption{Client skeleton of SRP-3, which provides CPSA a starting point for analyzing SRP-3 from the client's perspective.}
    \label{fig:model_SRP-3_client_skeleton}
\end{figure}


\begin{figure}
    \input{figures/server-skeleton}
    \caption{Server skeleton of SRP-3, which provides CPSA a starting point for analyzing SRP-3 from the server's perspective.}
    \label{fig:model_SRP-3_server_skeleton}
\end{figure}


\begin{figure}
    \input{figures/listener-x}  
    \caption{Client skeleton of SRP-3 with listener for the value~$x$, which provides CPSA a starting point for analyzing SRP-3 from the client's perspective. The listener role helps CPSA determine whether an execution of SRP-3 can leak the value~$x$.}
    \label{fig:model_SRP-3_listener_x}
\end{figure}


\begin{figure}
   \input{figures/listener-v}   
    \caption{Server skeleton of SRP-3 with listener for the value~$v$, which provides CPSA a starting point for analyzing SRP-3 from the server's perspective. The listener role helps CPSA determine whether an execution of SRP-3 can leak the value~$v$.}
    \label{fig:model_SRP-3_listener_v}
\end{figure}


 \begin{figure}
 	\input{figures/input-malicious-server}  
    \caption{Modeling a malicious server in CPSA.  We define the {malserver} 
   role to behave like a client while having access to the legitimate server's initialized variables.
   The associated skeleton provides CPSA a starting point for analyzing the malicious server attack from the perspective of the malicious server.}
    \label{fig:model_malserver}
\end{figure}




\end{document}